\documentstyle[12pt]{article}

\newcommand{\be}{\begin{equation}}
\newcommand{\ee}{\end{equation}}
\begin{document}
\begin{titlepage}
\vspace{1cm}
\begin{flushright}
December 8, 1999 \\
hep-th/9912050
\end{flushright}
\begin{center}
{\Large \bf The Origin of Chiral Anomaly\\
and the Noncommutative Geometry}\\
\vspace{1cm}
P. Pre\v{s}najder \\
Department of Theoretical Physics, Comenius University \\
Mlynsk\'{a} dolina, SK-84215 Bratislava, Slovakia \\[10pt]
\vspace{1cm}
\begin{abstract}
We describe scalar and spinor fields on a noncommutative sphere starting from
canonical realizations of the enveloping algebra ${\cal A}={\cal U}(u(2))$. The
gauge extension of a free spinor model, the Schwinger model on a noncommutative
sphere, is defined and the model is quantized. The noncommutative version of the
model contains only a finite number of dynamical modes and is non-perturbatively
UV-regular. An exact expression for the chiral anomaly is found. In the
commutative limit the standard formula is recovered.
\end{abstract}
\end{center}
\end{titlepage}

\section{Introduction}
The basic notions of the noncommutative geometry were developed 
in \cite{Con1}-\cite{Con3}, in the form of the matrix geometry they apeared in
\cite{DV}, \cite{DKM}. The essence of this approach consists in reformulating
first the geometry in terms of commutative algebras and modules of smooth
functions, and then generalizing them to their noncommutative analogs. The
notion of the space, the continuum of points, is lost, and this is expected to
lead to an UV-regular quantum field theory.

One of the simplest models for a noncommutative manifold is the
noncommutative (fuzzy) sphere, see \cite{Ber}-\cite{GP1}. Simple field
theoretical models containing scalar and spinor fields on a fuzzy sphere are
described in \cite{GP2}-\cite{GKP3}, an alternative form of the Dirac operator
was proposed in \cite{CW1}-\cite{CW2}. The issue of gauge field was investigated
in \cite{K}. A systematic approach to other structures on a Fuzzy sphere, like
monopoles and instantons, was proposed recently in \cite{Bal1}-\cite{Bal2}. All
models in question posses only finite number
of modes and are UV-regular.

Our aim is to apply the ideas of a noncommutative geometry to the
Schwinger model (QED in 2D) on a fuzzy sphere. The commutative version was
analyzed in detail in \cite{Jay}. Its noncommutative matrix version was
proposed in \cite{GM}. In \cite{K2} a supersymmetric Schwinger model was
described possessing finite number of modes. In \cite{GP4} we proposed an
approach going beyond matrix models. Here we extend these investigations. 

The construction of fuzzy field-theoretical models consists in two steps:

1) Fuzzy kinematics: We describe first scalar and spinor fields on a
noncommutative (fuzzy) shere starting from a canonical realization of the
enveloping algebra ${\cal A}={\cal U}(u(2))$, working within the set of all
finite dimensional fuzzy realizations of fields. Using a noncommutative version
of a prepotential formalis we perform straightforwardly the gauge extension of
the model - the Schwinger model. It contains fuzzy fields: the spinor fields
${\bar \Psi}$, $\Psi$, the dynamical gauge field $\sigma$
and a fuzzy gauge field $\lambda$ corresponding to the pure gauge. To gaurantee
the local gauge symmetry the field $\lambda$ should be assumed within all its
fuzzy realizations.

2) Fuzzy dynamics: We define the gauge invariant fuzzy Schwinger model action
depending on a finite number of modes of dynamical fields ${\bar \Psi}$, $\Psi$
and $\sigma$. We quantize the Schwinger model within functional integral
formalism integrating over dynamical modes. The truncation of modes cannot by
performed for the the pure gauge field $\lambda$, however the mean values of
gauge invariant functionals are $\lambda$-independent: the intengration over
finite number of dynamical modes gives the same result for any fixed $\lambda$.
This guarantees the nonperturbative UV-regularity of the model and allows us to
calculate excatly the noncommutative chiral anomaly. Using an explicit
$*$-product formula on a sphere we recover in the commutative limit the standard
result.

In Section 2 we describe the fuzzy kinematics of the Schwinger model. In Section
3 we formulate its quantum fuzzy dynamics and calculate exactly fuzzy chiral
anomaly. Last Section 4 contains concluding remarks. 

\section{Fields and gauge invariance}
\subsection{The scalar field}
We describe the scalar field on a sphere $S^2$ in the
$SU(2)$-invariant formulation, see \cite{Jay}. The sphere is interpreted as the
Hopf fibration
\[
S^3 =\{ \chi \in {\bf C}^2;\ \chi^+\chi =\rho \} \ \to \ S^2
=\{ x=(x_1 ,x_2 ,x_3 )\in {\bf R}^3 \}
\]
with $x_i =\frac{1}{2} \chi^+\sigma_i \chi ,i =1,2,3$. In this approach the
fields are functions of complex variables $\chi_\alpha$, $\chi^*_\alpha$,
$\alpha =1,2$. The space ${\cal H}_{k}$ of fields with a topological winding
number $2k\in {\bf Z}$ is defined as the space of analytical functions
\be
\Phi \ =\ \sum a_{mn} \chi^{*m} \chi^n \ ,\ |n|-|m|=2k\ .
\ee
We use the multi-index notation: if $n=(n_1 ,n_2)$ then $\chi^n =\chi^{n_1} \chi^{n_2}$, $|n|=n_1 +n_2 $, $n!=n_1 !n_2 !$, etc. Obviously, it holds ${\cal H}^*_k ={\cal H}_{-k}$ and ${\cal H}_k {\cal H}_l \subset {\cal H}_{k+l}$. The space ${\cal H}_0$ of functions constant on a fiber can be identified with algebra of functions on $S^2$. All spaces ${\cal H}_k$ are ${\cal H}_0$-modules. On ${\cal H}_k$ we introduce the scalar product 
\be
(\Phi_1 , \Phi_2 )\ =\ \int d\mu \Phi^*_1 \Phi_2 \ ,\
d\mu =(2\pi \rho )^{-1} d^3 x \delta (x^2_i - \rho^2 )\ .
\ee

The Poisson structure on ${\bf C}^2$ is defined by elementary brackets 
\be
\{ \chi_\alpha ,\chi_\beta \} \ =\ \{ \chi^*_\alpha ,\chi^*_\beta \} \ =\ 0\ ,\ \{ \chi_\alpha ,\chi^*_\beta \} \ =\ -i\delta_{\alpha \beta} \ ,\ \alpha ,\beta =1,2\ .
\ee
A Poisson bracket realizations of the $u(2)$ algebra is then obtained by choosing the basis:
\be
x_i \ =\ \frac{1}{2} \chi^+ \sigma_i \chi \ \ i=1,2,3\ ,\ r\ =\
\chi^+ \chi \ .
\ee
The functions $x_i$ generate $su(2)$, and $r$ is a central element
extending it to the $u(2)$ algebra. In ${\cal H}_k$ the $u(2)$ algebra is realized as the adjoint Poisson algebra:
\be
X_i \Phi \ =\ i\{ x_i ,\Phi \} \ ,i =1,2,3 \ ,\ R \Phi \ =\ i\{r,\Phi \} \ .
\ee
This allows to construct Wigner $D$-functions in a standard way: (i) the lowest weight is given by
\be
D^j_{k,-j} \ =\ [(2j+1)!/(j+k)!(j-k)!]^{1/2} r^{-j} \chi^{*j+k}_2 
\chi^{j-k}_1 \ ,
\ee
(ii) for a given $j$ all other $D^j_{km}$ are obtained by a repeated
action of $X_+$,
\be
D^j_{km} \ =\ [(j+m)!/(j-m)!(2j)!]^{1/2} \ X^{j-m}_+ D^j_{k,-j} \ .
\ee
It holdst $X_0 D^j_{k,-j} =-jD^j_{k,-j}$ and $X_- D^j_{k,-j} =0$ (here $X_0 = X_3$, $X_\pm = X_1 \pm iX_2$). The functions $D^j_{km}$, $|m|\le j$, $j = |k|, |k|+1, \dots$, form an orthonormal basis in ${\cal H}_k$ with respect to the scalar product (2).

In the noncommutative version we "quantize" the Poisson structure given above. We replace the commuting complex parameters $\chi_{\alpha}$, $\chi^*_{\alpha}$, by annihilation and creation operators ${\hat \chi}_{\alpha}$, ${\hat \chi}^*_{\alpha}$ satisfying commutation relations
\be
[{\hat \chi}_\alpha ,{\hat \chi}_\beta ]\ =\ [{\hat 
\chi}^*_\alpha ,{\hat \chi}^*_\beta ]\ =\ 0\ ,\ [{\hat 
\chi}_\alpha ,{\hat \chi}^*_\beta ]\ = \ \delta_{\alpha \beta} \ , \alpha = 1,2 \ .
\ee
They act in the auxiliary Fock space ${\cal F}=\{ |n\rangle =\frac{1}{\sqrt{n!}} {\hat \chi}^{*n} |0\rangle ,n=(n_1 ,n_2) \}$.

The space ${\tilde {\cal H}}_k$ of fields with topological winding number $2k$ is formed by operators  of the form
\be
{\tilde \Phi}\ =\ \sum a_{mn} {\hat \chi}^{*m} {\hat \chi}^n \ ,\ |n|-|m|=2k\ ,
\ee
defined on the invariant domain ${\cal F}_f =\{ \sum \alpha_n |n\rangle$ - finite sum$\}$. It holds ${\tilde {\cal H}}^*_k ={\tilde{\cal H}}_{-k}$ and ${\tilde {\cal H}}_k {\tilde{\cal H}}_l \subset {\tilde{\cal H}}_{k+l}$. The space ${\tilde{\cal H}}_0$ itself is a faithful canonical realization of the enveloping algebra            
${\cal A}={\cal U}(u(2))$ generated by
\be
{\hat x}_i \ =\ \frac{1}{2} {\hat \chi}^+ \sigma_i {\hat \chi}\ \ 
,\ \ i =1,2,3\ ,\ {\hat r}\ =\ {\hat \chi}^+ {\hat \chi} \ .
\ee
The subspace ${\cal F}_N =\{ |n\rangle \in {\cal F},|n|=N\}$ is a carrier space of the unitary irreducible $SU(2)$ representation corresponding to the spin $s=\frac{N}{2}$. The $u(2)$ action is realized in ${\tilde {\cal H}}_k$ as the adjoint commutator action
\be
{\hat X}_i {\hat \Phi}\ =\ i[{\hat x}_i ,{\hat \Phi}]\ ,i =1,2,3 \ ,\ {\hat R} {\hat \Phi} \ =\ i[\hat r,\Phi ]\ .
\ee
The fuzzy analogs of the Wigner $D$-functions can be constructed analogously as in the commutative case: (i) the lowest weight is 
\be
{\hat D}^j_{k,-j} \ =\ [(2j+1)!/(j+k)!(j-k)!]^{1/2} {\hat \chi}^{*j+k}_2 N_j ({\hat r}) {\hat \chi^{j-k}}_1 \ ,
\ee
(ii) the other $D^j_{km}$ are obtained by a repeated action of ${\hat X}_+$,
\be
{\hat D}^j_{km} \ =\ [(j+m)!/(j-m)!(2j)!]^{1/2} \ {\hat X}^{j-m}_+ {\hat D}^j_{k,-j} \ .
\ee
The factor $N_j ({\hat r})=[({\hat r}+j+1){\hat r}!/({\hat r}+2j+1)!]^{-1/2}$ is diagonal, it guarantees that the restricted $D$-functions ${\hat D}^{Jj}_{km} \equiv {\hat D}^{MN}_{jm} :={\hat D}^j_{km} |_{{\cal F}_N}$, $|m|\le j$, $j = |k|, |k|+1, \dots ,J$, form an orthonormal basis in the space of linear mappings
\[
{\hat{\cal H}}^J_k \equiv {\hat{\cal H}}_{MN} \ =\ \{
{\cal F}_N\ \to \ {\cal F}_N\ ;\ M=J+k, N=J-k\}\ ,
\]
endowed with the scalar product
\be
({\hat \Phi}_1 ,{\hat \Phi}_2 )^J_k \ =\ \frac{1}{J+1} {\rm Tr}({\hat \Phi}^*_1 {\hat \Phi}_2 )\ .
\ee
The relations $M=J+k, N=J-k$ among $M,N,J$ and $k$ we shall assume in what follows. 

Any operator ${\hat \Phi}\in {\tilde{\cal H}}_k$ can be expanded as
\be
{\hat \Phi}\ =\ \sum_{j=|k|}^\infty \sum_{|m|\le j} a^j_{km} ({\hat r}){\hat D}^j_{km} \ .
\ee
The relation ${\tilde {\cal H}}_k {\tilde{\cal H}}_l \subset {\tilde{\cal H}}_{k+l}$ takes for basis functions the form
\be
{\hat D}^{j'}_{k'm'}{\hat D}^{j''}_{k''m''}\ =\ \sum_j
{\tilde C}^{j'j''j}_{k'k''k} ({\hat r}) C^{j'j''j}_{m'm''m}{\hat D}^j_{km} \ .
\ee
Here $C^{j'j''j}_{m'm''m} \sim \delta_{m'+m'',m}$ are standard Clebsh-Gordon coefficients. The deformed coefficients ${\tilde C}^{j'j''j}_{k'k''k} ({\hat r})\sim \delta_{k'+k'',k}$ are square roots of a rational functions of ${\hat r}$, they possess the asymptotic expansion
\be
{\tilde C}^{j'j''j}_{k'k''k} ({\hat r})\ =\ C^{j'j''j}_{k'k''k} \ +\ \sum_{s=1}^{\infty} {\hat r}^{-s} C^{j'j''j}_{k'k''k} (n)\ =\
C^{j'j''j}_{k'k''k} \ +\ o({\hat r}^{-1})\ .
\ee

By ${\hat{\cal H}}_k$ we denote the subset of operators from ${\tilde {\cal H}}_k$ with the expansion coefficients
$a^j_{km} ({\hat r})$ possessing an asymptotic expansion
\be
a^j_{km} ({\hat r})\ =\ a^j_{km} \ +\ \sum_{s=1}^{\infty} a^j_{km} {\hat r}^{-s} \ .
\ee
For ${\hat{\cal H}}_k$ hold basic relations ${\hat{\cal H}}^*_k ={\hat{\cal H}}_{-k}$ and ${\hat{\cal H}}_k {\hat{\cal H}}_l \subset {\hat{\cal H}}_{k+l}$. Restricting the domain ${\cal F}_f$ to ${\cal F}_N$ one obtains the relations for the restricted spaces ${\hat{\cal H}}_{MN} \equiv {\hat{\cal H}}^J_k$:
\[
{\hat{\cal H}}^*_{MN} ={\hat{\cal H}}_{NM} \ ,\ {\hat{\cal H}}_{ML} {\hat{\cal H}}_{LN} \subset {\hat{\cal H}}_{MN} \ .
\]
Putting $M=J+k=J'+k'$, $L=J''+k''=J'-k'$ and $N=J-k=J''-k''$, the preceding relations can be rewritten as 
\[
{\hat{\cal H}}^{*J}_k ={\hat{\cal H}}^J_{-k} \ ,\ {\hat{\cal H}}^{J'}_{k'} {\hat{\cal H}}^{J''}_{k''} \subset {\hat{\cal H}}^J_k \ .
\]
If we take $k',k''$ and $J$ as independent then $J'=J+k''$, $J''=J-k'$ and $k=k'+k''$. The product formula for the restricted basis elements reads 
\be
{\hat D}^{J'j'}_{k'm'}{\hat D}^{J''j''}_{k''m''}\ =\ \sum_j
{\tilde C}^{j'j''j}_{MLN} C^{j'j''j}_{m'm''m}{\hat D}^{Jj}_{km} \ , 
\ee
where $M=J+k$, $L=J-k'+k''$ and $N=J-k$. The nonvanishing deformed coefficients are given by the formula
\be
{\tilde C}^{j',j'',j}_{MLN} \ =\ \frac{({\hat D}^{J,j}_{k'+k'',j'-j''},{\hat D}^{J',j'}_{k',j'}{\hat D}^{J'',j''}_{k'',j'})^J_k}{C^{j',j'',j}_{j',-j'',j'-j''}}\ =\ C^{j',j'',j}_{k',k'',k'+k''} \ +\ o(J^{-1}) \ .
\ee
They are square roots of rational functions in $J$, they break the commutativity (in general, they are not symmetrical with respect to primed and double primed indices), and they restrict the summation to $|k|\le j\le \min (j'+j'',J)$. 

To any field on a standard sphere (the "classical/commutative observable")
\be
\Phi_k \ =\ \sum_{j=|k|}^\infty \sum_{|m|\le j} a^j_{km} D^j_{km} 
\ \in \ {\cal H}_k \ ,
\ee
we assign the operator (the "quantum/fuzzy observable")
\be
{\hat \Phi}_k \ =\ \sum_{j=|k|}^\infty \sum_{|m|\le j} 
a^j_{km} ({\hat r}){\hat D}^j_{km} \ \in \ {\hat{\cal H}}_k\ ,
\ee
with $a^j_{km} ({\hat r})=a^j_{km} +o({\hat r}^{-1})$. Various choices of the $o({\hat r}^{-1})$ term corresponds to various fuzzy(fication) rules (in quantum mechanics they would correspond to various $\hbar$ terms generated by different operator orderings).
Restricting the domain ${\cal F}_f$ to ${\cal F}_N$ we obtain for any ${\hat \Phi}_k$ the infinite sequence of restrictions 
\be
{\hat \Phi}^J_k \equiv {\hat \Phi}_{MN} \ =\ {\hat \Phi} |_{{\cal F}_N} \ =\ \sum_{j=|k|}^J \sum_{|m|\le j} a^{Jj}_{km} {\hat D}^{Jj}_{km} \in {\hat{\cal H}}^J_k \ ,
\ee
with $a^{Jj}_{km} =a^j_{km}(J+k)= a^j_{km} +o(J^{-1} )$ (the truncation of the summation follows from ${\hat D}^j_{km} |_{{\cal F}_N} =0$ for $J>N+k=J$).

{\it Note 1}: Let $\Phi \in {\cal H}_k$, $A,B\in {\cal H}_0$. According to the fuzzy rules (21)-(22) we assign to any linear transformation $\Phi \to A\Phi B$ (=$AB\Phi$) in ${\cal H}_k$ the  mapping in the fuzzy space ${\hat{\cal H}}_k$ defined as
\be
{\hat \Phi} \ \to \ {\hat A}_L {\hat B}_R {\hat \Phi}\ :=\ {\hat A}{\hat \Phi}{\hat B}\ .
\ee
Restricting the invariant domain ${\cal F}_f$ to ${\cal F}_N$ we obtain the linear transformation ${\hat{\cal H}}_{MN} \equiv {\hat{\cal H}}^J_k \to {\hat{\cal H}}^J_k \equiv {\hat{\cal H}}_{MN}$, ($M=J+k$, $N=J-k$), given by the formula
\be
{\hat \Phi}_{MN} \ \to \ {\hat A}_M {\hat \Phi}_{MN} {\hat B}_N \ ,
\ee
(here we put simply ${\hat A}_M ,{\hat B}_N$ instead of ${\hat A}_{MM} ,{\hat B}_{NN}$). The determinant of this transformation is given by the well-known formula,
\be
{\rm det}^J_k {\hat A}_L {\hat B}_R \ =\ ({\rm det}_M A_M )^{N+1} \ ({\rm det}_N B_N )^{M+1} \ .
\ee

{\it Note 2}: Let us define the commutative limit as the inverse mapping to the fuzzy rules: to any class of operators of the form
\be
{\hat \Phi}\ =\ \sum_{j=|k|}^\infty \sum_{|m|\le j} 
[a^j_{km} +o({\hat r}^{-1})]{\hat D}^j_{km} \ ,
\ee
we assign the field on a commutative sphere
\be
[{\hat \Phi}]\ =\ \sum_{j=|k|}^\infty \sum_{|m|\le j} 
a^j_{km} D^j_{km} \ .
\ee
Due to the product formulas (16), (17) it holds $[{\hat D}^{j'}_{k'm'}{\hat D}^{
j''}_{k''m''}] =[{\hat D}^{j'}_{k'm'}]$ $[{\hat D}^{j''}_{k''m''}]$. Similarly,
(19), (20) guarantees that $[{\hat D}^{Jj'}_{k'm'}{\hat D}^{J''j''}_{k''m''}]=[
{\hat D}^{J'j'}_{k'm'}][{\hat D}^{J''j''}_{k''m''}]$, $J'=J+k''$, $J''=J-k'$,
provided that $J\ge j'+j''$. Consequently, algebraic relations among fuzzy
polynomials are, up to $o(J^{-1} )$ corrections, the same as their commutative analogs. For example, for polynomial operators ${\hat \Phi}\in {\hat{\cal H}}_k$, ${\hat A},{\hat B}\in {\hat{\cal H}}_0$, the relation ${\hat \Phi}_{MN} \to {\hat A}_M {\hat \Phi}_{MN} {\hat B}_N$ reduces in the commutative limit ($J=\frac{1}{2}(M+N)\to \infty$, $k=\frac{1}{2}(M
-N)$-fixed) to the commutative relation $\Phi \to A\Phi B$ (this can be extended
to the case when Wigner expansion coefficients of all operators in question are
rapidly decreasing). This is just the "correspondence principle" on a sphere
\cite{Ber} (in more general context see \cite{H1}, \cite{H2}). However, for a
more complex objects, like the determinant (26), the $o(J^{-1})$ corrections can
accumulate, and an "anomaly" can apear in the commutative limit.

\subsection{The spinor field}
Spinor fields in the standard (commutative) $SU(2)$-invariant formalism are 2-component functions
\be
\Psi \ =\ \left( \begin{array}{c} \Psi_1 \\ \Psi_2 \end{array} \right) \ =\ f \chi_+ +g\chi_- \ ,\ \ \chi_+ =\left( \begin{array}{c} \chi_1 \\ \chi_2 \end{array} \right) \ ,\ \chi_- =\left( \begin{array}{c} \chi^*_2 \\ -\chi^*_1 \end{array} \right) \ ,
\ee
with $\Psi_1 =\Psi_1 (\chi^* ,\chi )$, $\Psi_2 =\Psi_2 (\chi^* ,\chi )$, $f=f (\chi^* ,\chi )$ and $g=g(\chi^* ,\chi )$ being functions in complex variables $\chi^*_\alpha ,\chi_\alpha $, $\alpha =1,2$. The space ${\cal S}_k$ of spinor fields with winding number $2k\in {\bf Z}$ is formed by fields with $\Psi_1 , \Psi_2 \in {\cal H}_k$, or equivalently, with $f\in {\cal H}_{k+1/2}$ and $g\in {\cal H}_{k-1/2}$. 

In order to define the Dirac operator we introduce an orthonormal frame on $S^2 =\{ \vec{x}\in{\bf R}^3 ;\vec{x}^2 =\rho^2 \}$ as follows:
\be
\vec{y}_0 =\rho^{-1} \vec{x} = \rho^{-1}\chi^*_\alpha \vec{\sigma}_{\alpha\beta} \chi_\beta\ ,\ \vec{y}_+ =\vec{y}^*_- = \rho^{-1}\chi^*_\alpha \vec{\sigma}_{\alpha\beta} \varepsilon_{\beta\gamma}\chi^*_\gamma \ .
\ee
Here $\vec{y}_0$ is an unit vector perpendicular to $S^2$, and $\vec{y}_\pm$ are complex normalized vectors tangential to $S^2$, ($\varepsilon_{\alpha\beta}$ - antisymmetric, $\varepsilon_{12} =1$). The projections of the spin angular momentum $\frac{1}{2}\vec{\sigma}$ are
\[
\frac{1}{2}\Gamma_\pm \ =\ \frac{1}{2}\vec{y}_\pm .\vec{\sigma} \ ,\ \frac{1}{2}\Gamma_0 \ =\ \frac{1}{2}\vec{y}_0.\vec{\sigma} \ . 
\]
The Clifford algebra relations are satisfied: $ \Gamma^2_\pm =0$, $\Gamma_+\Gamma_- +\Gamma_-\Gamma_+ =1$, $ \Gamma_+\Gamma_- -\Gamma_-\Gamma_+ =2\Gamma_0$. Similarly, the projection of the orbital angular momemum $\vec{X}$ (the $SU(2)$ right-invariant vector fields defined in (5)) are given as 
\[
R_\pm \ =\ \vec{y}_\pm .\vec{X}\equiv K_\pm \ ,\ R_0 \ =\ \vec{y}_0.\vec{X}\equiv K_0 \ . 
\]
Here,
\be
K_+ \ =\ -\varepsilon_{\alpha\beta} \chi^*_\alpha \partial_{\chi_\beta} \ ,\ K_- \ =\ \varepsilon_{\alpha\beta}  \chi_\alpha\partial_{\chi^*_\beta} \ ,\ K_0 \ =\ \frac{1}{2}(\chi^*_\alpha \partial_{\chi^*_\alpha} -\chi_\alpha \partial_{\chi_\alpha}) \ , 
\ee
are the $SU(2)$ left-invariant vector fields. The free Dirac operator is given as
\be
D_0 \ \equiv \ I_+ \Gamma_- \ +\ I_- \Gamma_+ \ =\ K_+ \Gamma_- \ +\ K_- \Gamma_+\ +\ 1\ ,
\ee
where, 
\[
I_\pm =K_\pm +\frac{1}{2}\Gamma_\pm \ ,\ I_0 =K_0 +\frac{1}{2}\Gamma_0 \ , 
\]
are projectionts of the total angular momenum $\vec{J}=\vec{X}+\frac{1}{2}\vec{\Gamma}$ into basis (30). The Dirac operator $D_0$ anticommutes with the chirality operator $\Gamma_0$. Since $\Gamma_0 \chi_\pm =\pm \chi_\pm$, $f$ and $g$ are chiral components of $\Psi$ to the chiralities +1 and -1 respectively.
The operator $D_0$ already contains the $2k$-monopole field strenght $F_0$. This can be identified with the term in $D^2_0$ proportional to $\Gamma_0$. It can be found straightforwardly: $F_0 =\frac{1}{2}[R_+ ,R_- ]\Gamma_0 =K_0 \Gamma_0$.
The operator $R_0 =K_0$ takes in ${\cal S}_k$ constant value $k$. Obviously, $F_0 |_{{\cal S}_k} =k\Gamma_0$ is the monopole field in question. 

Standard formulas are obtained by replacing in the formulas above $\chi^*_\alpha$, $\chi_\alpha$ by arbitrary $S^3 \to S^2$ sections $\chi^*_\alpha (\vec{x})$, $\chi_\alpha (\vec{x})$. Below we shall use an equivalent $SU(2)$-invariant supersymmetric picture which is appropiate for the noncommutative generalization (see \cite{GKP2} - \cite{GKP3}).

In this formalism the fields are functions on ${\bf C}^{2|1}$ in complex variables $\chi_\alpha$, $\chi^*_\alpha$, $\alpha =1,2$, and one anticommuting (Grassmannian) pair $a, a^*$. The space of spinor fields with given winding number $2k$ is defined as
\[
{\cal S}_k \ =\ {\cal H}_{k+\frac{1}{2}} a\ \oplus
\ {\cal H}_{k-\frac{1}{2}} a^* \ =\ \{ \Psi = f a + g a^* ;f\in
{\cal H}_{k+\frac{1}{2}} ,g\in {\cal H}_{k-\frac{1}{2}} \} \ .
\]
In this approach the analogs of the spin angular momentum projections are 
\[
\frac{1}{2}\Gamma_+ =\frac{1}{2} a\partial_{a^*}\ ,\ \frac{1}{2}\Gamma_- =\frac{1}{2} a^*\partial_a \ ,\ \frac{1}{2}\Gamma_0 =\frac{1}{2} a\partial_a -a^*\partial_{a^*} \ .
\]
The Clifford algebra relations are satisfied. The operators $P_+ =a\partial_a$ and $P_- =a^*\partial_{a^*}$ are projection onto spinor subspaces with chiralities $+1$ and $-1$ respectively. Similarly, the analogs of the total angular momentum projections are
\[
I_\pm \ =\ K_\pm \ ,\ I_0 \ =\ K_0 \ ,
\]
so the analogs of the orbital angular momentum projections are given as
\[
R_\pm \ =\ K_\pm -\frac{1}{2}\Gamma_\pm \ ,\ I_0 \ =\ K_0 -\frac{1}{2}\Gamma_0\ .
\]

In ${\cal S}_{k}$ the free Dirac operator is given by the formula (see (32)):
\be
D_0 \ \equiv \ I_+ \Gamma_- \ +\ I_- \Gamma_+ \ =\ K_+ \Gamma_- \ +\ K_+ \Gamma_-\ .
\ee
It anticommutes with the chirality operator $\Gamma_0 =P_+ -P_-$. The monopole field is: $F_0 =\frac{1}{2}[R_+ ,R_- ]\Gamma_0 =K_0 \Gamma_0 -\frac{1}{2}$.

The charge conjugation ${\cal J}$ is defined as follows: ${\cal J}(fa+ga^* )=g^* a-f^* a^*$. Obviously, ${\cal J}:{\cal S}_{k} \to {\cal S}_{-k}$, and ${\cal J}^2 =-1$. The inner product in ${\cal S }_k$ we define as
\be
(\Psi_1 , \Psi_2 )\ =\ \int d\nu {\cal J} \Psi_1 \Psi_2
\ =\ \int d\mu (f^*_1 f_2 + g^*_1 g_2 )\ ,
\ee
where $d\nu =(16\pi^2 \rho )^{-1} d^2 \chi^* d^2 \chi dada^* \delta
(\chi^*_\alpha \chi_\alpha + a^* a - \rho )$. Any spinor field from ${\cal S}_k$ can be expanded as
\be
\Psi \ =\ \sum_{j=|k|-1/2}^{\infty} \sum_{|m|\le j} [a^{j+}_{km}
D^j_{k-1/2,m} a\ +\ a^{j-}_{km} D^j_{k+1/2,m} a^* ] \ .
\ee
The first term with $j=|k|-\frac{1}{2}$ appears only for $k>0$ with 
$\Gamma =-1$, or for $k<0$ with ($\Gamma =+1$).
 
In the space of functions on ${\bf C}^{2|1}$ a graded Poisson structure can be introduced by postulating elementary graded brackets
\be
\{ \chi_\alpha ,\chi^*_\beta \} \ =\ -i\delta_{\alpha \beta} \ ,\
\{ a, a^* \} \ =\ -i\ ,
\ee
(all other elementary brackets vanish). The operators $K_\pm$ and $\Gamma_\pm$ can be expressed in terms of Poisson brackets as follows
\[
K_- \ =\ -i\varepsilon_{\alpha \beta} \chi_\alpha \{ \chi_\beta ,.\} \ , \ K_+ \ =\ i\varepsilon_{\alpha \beta} \chi^*_\alpha \{ \chi^*_\beta ,.\} \ ,
\]
\be
\Gamma_- \ =\ ia \{ a,.\} \ ,\ \Gamma_+ \ =\ ia^* \{ a^* ,.\} \ .
\ee

{\it Note 1}: The $u(2|1)$ Poisson bracket superalgebra is realized by choosing the basis
\[
x_i \ =\ \frac{1}{2} \chi^+ \sigma_{i} \chi \ ,\ i=1,2,3 \ ,\
b\ =\ \chi^+ \chi \ +\ 2 a^* a \ ,\
\]
\be
v_\alpha \ =\ \chi_\alpha a^* \ ,\ {\bar v}_\alpha \ =\
\varepsilon_{\alpha \beta} \chi^*_\beta a\ ,\ \alpha =1,2 \ ,\
s \ =\ \chi^+ \chi \ +\ a^* a \ .
\ee
The operators $x_i$ and $b$ are even generators, and $v_\alpha ,{\bar
v}_\alpha$ the odd ones of the superalgebra $su(2|1)$; $s$ is a central element extending it to the $u(2|1)$ superalgebra. The adjoint Poisson bracket representation of the $u(2|1)$ superalgebra is realized in the space of superfields:
\[
X_i \Psi \ =\ i\{ x_i ,\Psi \} \ ,\ B \Psi \ =\ i\{ b,\Psi \} \ ,\
S \Psi \ =\ i\{ s,\Psi \} \
\]
\be
V_\alpha \Psi \ =\ \{ v_\alpha ,\Psi \} \ ,\
{\bar V}_\alpha \Psi \ =\ \{ {\bar v}_\alpha ,\Psi \} \ .
\ee
This action can be extended to the enveloping superalgebra ${\cal B}={\cal U}(u(2|1))={\cal B}^0 \oplus {\cal B}^1$. The spinor space ${\cal S}_k$ is invariant with respect to the adjoint action (39) generated by the even subalgebra ${\cal B}^0$. The free Dirac operator can be expressed as: $D_0 =\frac{1}{2} \varepsilon_{\alpha \beta} (V_\alpha V_\beta +{\bar V}_\alpha {\bar V}_\beta )$.

In the fuzzy case we quantize the graded Poisson structure (36): we replace the graded commuting variables $\chi_{\alpha} ,\chi^*_{\alpha}$ and $a$, $a^*$ by annihilation and creation operators ${\hat\chi}_{\alpha} ,{\hat \chi}^*_{\alpha}$ (bosonic) and ${\hat a},{\hat a}^*$ (fermionic) satisfying graded commutation relations 
\be
[{\hat \chi}_\alpha ,{\hat \chi}^*_\beta ]\ =\ \delta_{\alpha \beta} \ ,\
[ a, a^* ] \ =\ 1
\ee
(all other elementary brackets vanish). They act in the auxiliary Fock superspace $s{\cal F}=\{ |n\nu \rangle =(n!)^{-1/2}{\hat \chi}^{*n} {\hat a}^{*\nu} |0\rangle ={\cal F}\oplus {\hat a}^*{\cal F}$. The subspace of spinor operators with winding number $2k$ we define as the space
\[
{\tilde{\cal S}}_k \ =\ {\tilde{\cal H}}_{k+\frac{1}{2}} {\hat a}\ \oplus \ {\tilde{\cal H}}_{k-\frac{1}{2}} {\hat a}^* \ =\ \{ {\tilde \Psi} = {\tilde f}{\hat a} +{\tilde g}{\hat a}^* ;{\tilde f}\in
{\tilde{\cal H}}_{k+\frac{1}{2}} ,{\tilde g}\in {\tilde{\cal 
H}}_{k-\frac{1}{2}} \} \ 
\]
of odd mappings $s{\cal F}\to s{\cal F}$ defined on the invariant domain $s{\cal F}_f ={\cal F}_f \oplus {\cal F}_f {\hat a}^*$ (of finite linear combinations of states $|n,\nu \rangle$). The charge conjugation ${\cal J}$ is defined as follows: ${\cal J}{\tilde f}{\hat a}+{\tilde g}{\hat a}^* ={\tilde g}^* {\hat a}-{\tilde f}^* {\hat a}^*$. Obviously, ${\cal J}:{\tilde{\cal S}}_{k} \to {\tilde{\cal S}}_{-k}$, and ${\cal J}^2 =-1$.

The space ${\tilde{\cal S}}_{k}$ is a bi-module with respect to the left and right multiplications by the elements of the algebra ${\tilde{\cal B}}_0 ={\tilde{\cal H}}_0 \oplus {\hat a}^* {\hat a} {\tilde{\cal H}}_0$. Any element from ${\tilde{\cal S}}_k$ can be expanded as
\be
{\tilde \Psi}\ =\ \sum_{j=|k|-1/2}^{\infty} \sum_{|m|\le j} [ a^{j+}_{km} ({\hat r}){{\hat D}}^j_{k-1/2,m} a\ +\ a^{j-}_{km} ({\hat r}){{\hat D}}^j_{k+1/2,m} a^* ] \ .
\ee 

By ${\hat{\cal S}}_k ={\hat{\cal H}}_{k+ \frac{1}{2}} {\hat a}\oplus  {\hat{\cal H}}_{k-\frac{1}{2}} {\hat a}^*$ we denote the subspace of ${\tilde{\cal S}}_{k}$ with $a^{j\pm}_{km} ({\hat r})=a^{j\pm}_{km} +o({{\hat r}}^{-1})$. The space ${\hat{\cal S}}_{k}$ is a ${\hat {\cal B}}_0 ={\hat{\cal H}}_0 \oplus {\hat a}^* {\hat a} {\hat{\cal H}}_0$ bi-module. In ${\hat{\cal S}}_{k}$ the free Dirac operator is defined as follows (see (33)):
\be
D_0 \ =\ {\hat K}_+ {\hat \Gamma}_- \ +\ {\hat K}_+ {\hat \Gamma}_-\ ,
\ee
where
\[
{\hat K}_- \ =\ -\varepsilon_{\alpha \beta} {\hat \chi}_\alpha [{\hat \chi}_\beta ,.] \ ,\ {\hat K}_+ \ =\ -\varepsilon_{\alpha \beta} {\hat \chi}^*_\alpha [ {\hat \chi}^*_\beta ,.] \ ,
\]
\be 
{\hat \Gamma}_- \ =\ {\hat a}[{\hat a},.] \ ,\ {\hat \Gamma}_+ \ =\ {\hat a}^* [{\hat a}^* ,.] \ .
\ee
It anticommutes with the  chirality operator ${\hat \Gamma}_0 ={\hat P}_+ -{\hat P}_-$. Here ${\hat P}_+ ={\hat a}[{\hat a}^* ,.]$ and ${\hat P}_- ={\hat a}^* [{\hat a},.]$ are projectors onto subspaces with chiralities $+1$ and $-1$, respectively.

{\it Note 2}: The fuzzy analogos ${\hat x}_i ,{\hat v}_\alpha ,{\hat
{\bar v}}_\alpha$ and ${\hat s}$ of generators satisfying in $s{\cal F}$ the
$u(2|1)$ graded commutator relations are given by eqs. (38), similarly the
$u(2|1)$ adjoint action and the Dirac operator are given by eqs. (39) and (42)
(of course, all commuting parameters are replaced by annihilation and creation
operators and graded brackets $i\{ .,.\}$ by graded commutators $[.,.]$). The
subspace $s{\cal F}_N ={\cal F}_N \oplus {\cal F}_{N-1} {\hat a}^*$ is the
carier space of the atypical unitary irreducible representation of $su(2|1)$
superalgebra.

To any spinor field  
\be
\Psi_k \ = \sum_{j=|k|-1/2}^{\infty} \sum_{|m|\le j} [ a^{j+}_{km} D^j_{k-1/2,m} a\ +\ a^{j-}_{km} D^j_{k+1/2,m} a^* ]\ \in \ {\cal S}_k
\ ,
\ee
we assign the spinor operator
\be
{\hat \Psi}\ = \sum_{j=|k|-1/2}^{\infty} \sum_{|m|\le j} [{\hat a}^{j+}_{km} ({\hat r}){{\hat D}}^j_{k-1/2,m} {\hat a}\ +\ a^{j-}_{km} ({\hat r}){{\hat D}}^j_{k+1/2,m} {\hat a}^* ] \ \in \ {\hat {\cal S}}_k \ ,
\ee
defined on the domain $s{\cal F}_f ={\cal F}_f \oplus {\cal F}_f {\hat a}^*$. Restricting it to the subspace $s{\cal F}_N ={\cal F}_N \oplus {\cal F}_{N-1} {\hat a}^*$ we obtain an infinite set of restrictions 
\[
{\hat \Psi}^J_k  \equiv {\hat \Psi}_{MN} = {\hat \Psi}_k |_{s{\cal F}_N}
\]
\be
\ = \sum_{j=|k|-1/2}^{J-1/2} \sum_{|m|\le j} [a^{j+}_{km} ({\hat r}){{\hat D}}^{J-1/2,j}_{k-1/2,m} {\hat a}\ +\ a^{j-}_{km} ({\hat r}){{\hat D}}^{J-1/2,j}_{k+1/2,m} {\hat a}^* ]\ ,
\ee
belonging to the space ${\hat{\cal S}}^J_k ={\hat{\cal H}}^{J-1/2}_{k+\frac{1}{2}}{\hat a}\oplus {\hat{\cal H}}^{J-1/2}_{k-\frac{1}{2}}{\hat a}^*$ of odd elements in the superspace of mappings $s{\cal H}^J_k =\{ s{\cal F}_N \to s{\cal F}_M ,\ M=J+k, N=J-k \}$. In ${\hat{\cal S}}^J_k$ we introduce the inner product
\be
({\hat \Psi}_1 ,{\hat \Psi}_2 )^J_k \ =\ \frac{1}{2J} s{\rm Tr}_N [{\cal J}{\hat \Psi}_1 {\hat \Psi}_2 ]\ =\ \frac{1}{2J}({\rm Tr}_N {\hat f}^*_1 {\hat f}_2 +{\rm Tr}_{N-1} {\hat g}^*_1 {\hat g}_2 ) ,
\ee
where $s{\rm Tr}_N$ denotes the supertrace in the space of mappings
$s{\cal F}_N \to s{\cal F}_N$. In ${\hat{\cal S}}^J_k$ the spectrum of $D_0$ is the same but truncated as in the commutative case (see \cite{GKP2}).

\subsection{The gauge field}
In the commutative case it is convenient to introduce gauge fields within the prepotentential formalism. In this approach the full Dirac operator $D$ is given by the formula:
\be
D\ =\ D_0 \ +\ A\ ,\ A\ =\ A_+ \Gamma_- \ +\ A_+ \Gamma_-\ .
\ee
The chiral components $A_\pm = K_\pm (\sigma \mp i\lambda )$ of the gauge potential are expressed in terms of two real prepotentials $\lambda \in {\cal H}_0$ and $\sigma \in {\cal H}_0$: the 
$\lambda$-dependent term $i[D_0 ,\lambda ]$ of $A$
represents a pure gauge field (corresponding to an exact 1-form in the differential form approach), whereas the $\sigma$-dependent term $[D_0 ,\sigma ]$ represents a dynamical gauge field (corresponding a co-exact 1-form). We note that $D_0$ already contains the monopole gauge field (corresponding to a harmonic 1-form).

In the $SU(2)$-invariant formalism the full Dirac operator can be written in the form
\be
D\ =\ {\cal I}_+ \Gamma_- \ +\ {\cal I}_- \Gamma_+ \ =\ {\cal R}_+ \Gamma_- \ +\ {\cal R}_+ \Gamma_-\ +\ 1\ .
\ee
where ${\cal I}_+ =\Omega I_+ \Omega^{-1}$, ${\cal I}_- =\Omega^{*-1} I_+ \Omega^*$,  analogously ${\cal R}_+ =\Omega R_+ \Omega^{-1}$, ${\cal R}_- =\Omega^{*-1} R_+ \Omega^*$. Here, $\Omega =e^{i\lambda} e^{\sigma}$ is an invertible element from ${\cal H}_0$ acting on $\Psi$ as a left multiplicator. The field strenght $F$ can be identified with the term in $D^2$ proportional to $\Gamma_0$:
\be
F \ =\ \frac{1}{2} [{\cal R}_+ ,{\cal R}_- ]\Gamma_0 \ =\ (\frac{1}{2} [{\cal R}_+ ,{\cal R}_- ]-R_0 ) \Gamma_0 \ +\  R_0 \Gamma_0 \ .
\ee
It can be shown straightforwardly that $\frac{1}{2} [{\cal R}_+ ,{\cal R}_- ]-R_0 = \Delta \sigma$, where $\Delta$ is the Laplace operator on a sphere. 

Using the relations $I_\pm =K_\pm$ and $R_\pm =K_\pm -\frac{1}{2}\Gamma_\pm$ valid in the $SU(2)$-invariant supersymmetric formalis, the expressions for the full Dirac operator and the field strenght read
\be
D \ =\ {\cal K}_+ \Gamma_- \ +\ {\cal K}_+ \Gamma_-\ ,
\ee
and
\be
F \ =\ (\frac{1}{2} [{\cal K}_+ ,{\cal K}_- ]-K_0 )\Gamma_0 \ +\  R_0 \Gamma_0 \ ,
\ee
with
\be 
{\cal K}_+ \ =\ \Omega K_+ \Omega^{-1} \ ,\  {\cal K}_- \ =\ \Omega^{*-1} K_+ \Omega^* \ .
\ee

Any unitary element $\omega \in {\cal H}_0$ generates a local $U(1)$ gauge transformations of all fields in question. The spinor field $\Psi =fa+ga^* \in {\cal S}_k$, the conjugated spinor field ${\bar \Psi}={\bar f}a+{\bar g}a^* \in {\cal S}_{-k}$ and the gauge field $\Omega \in {\cal H}_0$ transforms as follows:
\be
\Psi \ \to \ \omega \Psi \ ,\ {\bar \Psi} \ \to \ {\bar \Psi}\omega^* \ ,\ \Omega \ \to \ \omega \Omega \ .
\ee
Obviously, $e^{i\lambda} \to \omega e^{i\lambda}$ and $e^{\sigma}$ is gauge invariant. Moreover, it can be easily seen that under gauge transformations: $D\to \omega D\omega^*$, $F\to \omega F\omega^* =F$.
Consequently, the spinor term ${\bar \Psi}iD\Psi \in {\cal H}_0$ in the Schwinger model Lagrangian and the gauge field term $F^2 \in s{\cal H}_0$ are both gauge invarant. 

In the fuzzy supersymmetric picture we use exactly the same formulas as above,
however all objects (variables, fields and spaces) should be replaced
by their noncommutative partners. The full Dirac operator is an operator the noncommutative spinor space ${\hat{\cal S}}_k$ defined as 
\be
D\ =\ {\hat{\cal K}}_+ {\hat \Gamma}_- \ +\ {\hat{\cal K}}_+ {\hat \Gamma}_-\ ,
\ee
where
\be 
{\hat{\cal K}}_+ \ =\ {\hat \Omega}{\hat K}_+ {\hat \Omega}^{-1} \ ,\ {\hat{\cal K}}_- \ =\ {\hat \Omega}^{*-1}{\hat K}_+ {\hat \Omega}^* \ .
\ee
The operators ${\hat K}_\pm$: have been defined in Section 2, ${\hat \Omega}$ is an arbitrary invertible element from ${\hat{\cal H}}_0$. The field strenght operator ${\hat F}:{\hat {\cal S}}_k \to {\hat{\cal S}}_k$ we define in analogy with (52) as follows: 
\be
{\hat F}\ =\ (\frac{1}{2} [{\hat{\cal K}}_+ ,{\hat{\cal K}}_- ]-{\hat K}_0 ){\hat \Gamma}_0 \ +\ {\hat R}_0 {\hat \Gamma}_0 \ ,
\ee

The local gauge transformations of all fields in question, the spinor field ${\hat \Psi}\in {\hat{\cal S}}_k$, the conjugated spinor field ${\hat{\bar \Psi}}\in {\hat {\cal S}}_{-k}$ and the gauge field ${\hat \Omega}\in {\hat{\cal H}}_0$ are generated by unitary elements ${\hat \omega}\in {\hat{\cal H}}_0$. The gauge transformations rules read:
\be
{\hat \Psi}\ \to \ {\hat \omega}{\hat \Psi}\ ,\ {\hat{\bar \Psi}}\ \to \ {\hat{\bar \Psi}}{\hat \omega}^* \ ,\ {\hat \Omega}\ \to \ {\hat \omega}{\hat \Omega}\ .
\ee
Thus, the right-radial operator $e^{\hat \sigma}$ in the polar decomposition of
 ${\hat \Omega}= e^{i{\hat \lambda}} e^{\hat \sigma}$ is gauge invariant.
 Moreover, it can be easily seen that under gauge transformations: $D\to {\hat
 \omega} D{\hat \omega}^*$, ${\hat F}\to {\hat \omega}{\hat F}{\hat \omega}^*$.
 Consequently, the fuzzy analog of the spinor term ${\hat{\bar \Psi}}iD\Psi \in
 {\hat {\cal H}}_0$ in the Schwinger model Lagrangian is gauge invariant,
 whereas the gauge term ${\hat F}^2 \in s{\hat{\cal H}}_0$ transforms
 homogenously: ${\hat F}^2 \to {\hat \omega}{\hat F}^2 {\hat \omega}^*$.

\section{Quantization and chiral anomaly}
\subsection{Field action and quantization}

In the commutative case the Schwinger model field action
\be
S[\Psi ,{\bar \Psi} ,\Omega ]\ =\ \frac{1}{4q^2_o} \int d\mu F^2 \
+\ \int d\nu {\bar \Psi} D \Psi \ ,
\ee
is gauge invariant ($q_o$ is interaction constant). Introducing new spinor fields by putting
\[
\Psi \ \to \ \Psi_0 \ =\ {\tilde \Omega}^* \Psi \ :=\ \Omega^* fa\ +\
\Omega^{-1} ga^* \ ,
\]
\be
{\bar \Psi} \ \to \ {\bar \Psi}_0 \ =\ {\bar \Psi} {\tilde \Omega} \ :=\ {\bar f}\Omega^{*-1} a\ +\ {\bar g}\Omega a^* \ .
\ee
we obtain an action 
\be
S[\Psi_0 ,{\bar \Psi}_0 ,\sigma ]\ =\ \frac{1}{4q^2_o} \int d\mu F^2 \ +\ \int d\nu {\bar \Psi}_0 D_0 \Psi_0 \ ,
\ee
from which the electromagnetic interaction is eliminated, it describes a system of non-interacting gauge and spinor fields. We stress, that this is a valid procedure for the classical (non-quantized) fields only. 

We quantize the Schwinger model within the functional integral approach.
The dynamical fields in question $\Xi =\{ \Psi ,{\bar \Psi},\sigma \}$ we expand as follows: 
\[
\Psi \ =\ \sum_{j=|k|-1/2}^{\infty} \sum_{|m|\le j} [a^{j+}_{km}
D^j_{k-1/2,m} a\ +\ a^{j-}_{km} D^j_{k+1/2,m} a^* ]\ \in \ {\cal S}_k \ ,
\]
\[
{\bar \Psi} \ =\ \sum_{j=|k|-1/2}^{\infty} \sum_{|m|\le j} [{\bar a}^{j+}_{-km}
D^j_{-k-1/2,m} a\ +\ {\bar a}^{j-}_{-km} D^j_{-k+1/2,m} a^* ]\ \in \ {\cal S}_{-k} \ ,
\]
\be
\sigma \ =\ \sum_{j=1}^{\infty} \sum_{|m|\leq j} b^j_m D^j_{0m} \ \in \ {\cal H}_0 \ ,\ b^j_{-m} = (-1)^m b^{j*}_m \ .
\ee
The pure gauge prepotential $\lambda$ is an arbitrary fixed real fuction from ${\cal H}_0$. The action $S[\Xi ]$, given in (59), is a function of Grassmannian parameters $a^{j\pm}_{km}$, ${\bar a}^{j\pm}_{-km}$ and of complex parameters $b^j_m$.

The quantum mean values of gauge invariant field functionals $P[\Xi ]$ are defined by the formula
\be
\langle P[\Xi ]\rangle \ =\ Z^{-1} \ \int (D\Xi )_k \ P[\Xi ] \ e^{-S[\Xi ]} \ ,\ Z \ =\ {\int (D\Xi )_k \ e^{-S[\Xi ]}} \ .
\ee
They do not depend on the choice of $\lambda$. The symbol $(D\Xi )_k$ denotes the formal infinite dimensional integration over all dynamical field configurations:
\[
(D\Xi )_k \sim \left( \prod_{j=1}^{\infty} \prod_{m=1}^j db^j_m \right)
\]
\be
\ .\ \left( \prod_{|m|\leq |k|-1/2} da^0_{km} d{\bar a}^0_{-km} \right)\left( \prod_{j=|k|+1/2}^{\infty} \prod_{|m|\leq j} da^{\pm j}_{km} d{\bar a}^{\pm j}_{-km} \right)  
\ee
(here $a^0_{km}$ and ${\bar a}^0_{-km}$ correspond to zero modes). The mean values are defined only formally, and a regularization procedure is needed.

We may try to solve the model by performing the same spinor fields transformation (60). However, in the quantum case we have to take into account the determinat of this transformation (which is inverse to the corresponding Jacobian appearing in the functional integral). Formally, it is given as
\be
{\rm det}{\tilde \Omega} {\rm det}{\tilde \Omega}^* \ =\ {\rm det}\Omega \
{\rm det}\Omega^* \ {\rm det}\Omega^{*-1} \ {\rm det}\Omega^{-1} \ .
\ee
Naively, it equals to 1, but one should take into account that in the commutative case all determinants on r.h.s. are singular. Regularizing them properly it can be shown that (see e.g. \cite{Jay}):
\be
{\rm det}{\tilde \Omega} {\rm det}{\tilde \Omega}^* \ \sim \exp \{
\Sigma [\sigma ]\} \ ,\  \Sigma [\sigma ]\ =\ 2q^2_0 \int d\mu (X_i
\sigma (\vec{x}))^2 \ .
\ee
Thus, there apears a {\it chiral anomaly}, the nontrivial quantum correction $\Sigma [\sigma ]$ to the bosonic action (it generates a mass for an effective free bosonic field).

In the noncommutative case the calculations are simple and straightforward. We quantize the model according to the general formula (63), however there are important differences:  

i) We take the dynamical fields $\Xi =\{ {\hat \Psi},{\hat{\bar \Psi}},{\hat \sigma} \}$ in the following form: 
\[
{\hat \Psi}\ =\ \sum_{j=|k|-1/2}^{\infty} \sum_{|m|\le j} [a^{j+}_{km}
{\hat D}^j_{k-1/2,m} a\ +\ a^{j-}_{km} {\hat D}^j_{k+1/2,m} a^* ]\ \in \ {\hat{\cal S}}_k \ ,
\]
\[
{\hat{\bar \Psi}}\ =\ \sum_{j=|k|-1/2}^{\infty} \sum_{|m|\le j} [{\bar a}^{j+}_{-km}
{\hat D}^j_{-k-1/2,m} a\ +\ {\bar a}^{j-}_{-km} {\hat D}^j_{-k+1/2,m} a^* ]\ \in \ {\hat{\cal S}}_{-k} \ ,
\]
\be
{\hat \sigma}\ =\ \sum_{j=1}^{\infty} \sum_{|m|\leq j} b^j_m {\hat D}^j_{0m} \ \in \ {\hat{\cal H}}_0 \ ,\ b^j_{-m} = (-1)^m b^{j*}_m \ .
\ee
We choose all expansion coefficients without $o({\hat r}^{-1})$ terms (this fixes the fuzzy rules for dynamical fields).The pure gauge prepotential ${\hat \lambda}$ is an arbitrary fixed hermitean operator from ${\hat{\cal H}}_0$ (this fixes the gauge).

ii) We implement a natural regularization induced by the noncommuative geometry. We achieve this by fixing $J$ and taking the Schwinger model field action in the form
\be
S^J_k [{\hat \Psi},{\hat{\bar \Psi}} ,{\hat \sigma}] \ =\ \frac{1}{2J} s{\rm Tr}_N [{\hat{\bar \Psi}}iD{\hat \Psi}]\ +\ \frac{1}{4[(J+1/2)^2 -k^2]e^2_o} s{\rm Tr}^J_k [{\hat F}^2 \Gamma_0 ]\ .
\ee
Here, $e_o$ is interaction constant specified below. In the first term $s{\rm Tr}_N$ denotes the supertrace in the space of mappings $s{\cal F}_N \to s{\cal F}_N$. In the second term $s{\rm Tr}^J_k$ denotes the supertrace in the space of mappings ${\hat{\cal S}}^J_k \to {\hat{\cal S}}^J_k$. The gauge degrees of freedom enters the action via operator ${\hat \Omega}=e^{i{\hat \lambda}} e^{i{\hat \sigma}}$. 

The definition (68) is an essential step. It implies that only a finite number of dynamical modes takes part in the calculation of the action: 

- the Grassmannian spinor modes $a^0_{km}$, ${\bar a}^0_{-km}$ (zero
modes) and $a^{j\pm}_{km}$, ${\bar a}^{j\pm}_{km}$ for $j=|k|+1/2,|k|+3/2,
\dots ,J-1/2$, $|m|\leq j$, and

- the bosonic modes $b^j_m$ with $j =1,\dots J+k$, $0\leq m\leq j$ (the coefficients with negative $m$ are given by $b^j_{-m} =(-1)^m b^{j*}_m$).

The quantum mean values are again defined by (63), however the measure $(D\Xi )^J_k$ contains only relevant dynamical modes:
\[
(D\Xi )^J_k =\left( \prod_{j=0}^{J+k} db^j_0 \prod_{m=1}^j db^j_m \right) 
\]
\be
\ .\ \left( \prod_{|m|\leq |k|-1/2} \frac{da^0_{km}}{\sqrt{J}} \frac{d{\bar a}^0_{-km}}{\sqrt{J}} \right) \left(\prod_{j=|k| +1/2}^{J-1/2} \prod_{|m|\leq j}\frac{da^{\pm j}_{km}}{\sqrt{J}} \frac{d{\bar a}^{\pm j}_{-km}}{\sqrt{J}}\right) .
\ee
Its dimension is finite, and consequently there are no UV divergencies. This allows to calculate straightforwardly various non-perturbative quantities.

As an example, let us consider the problem of chiral anomaly. Applying the transformation (60) in the functional integral, we have to take into account that the  spinor fields are restricted from ${\hat{\cal S}}_{\pm k}$ to ${\hat{\cal S}}^J_{\pm k}$. The restricted transformation (60) reads
\[
{\hat f}_{M,N-1} {\hat a}+{\hat g}_{M-1,N} {\hat a}^* \ \to \ {\hat \Omega}^*_M {\hat f}_{M,N-1} {\hat a}+{\hat \Omega}^{-1}_{M-1} {\hat g}{\hat a}^* \ ,
\]
\be
{\hat{\bar f}}_{N,M-1} {\hat a} +{\hat{\bar g}}_{N-1,M} {\hat a}^* \ \to \ {\hat{\bar f}}_{N,M-1} {\hat\Omega}^{*-1}_{M-1} {\hat a}+{\hat{\bar g}}_{N-1,M} {\hat \Omega}_M {\hat a}^* \ ,
\ee
(we put simply ${\hat \Omega}_M$ instead of ${\hat \Omega}_{MM}$). The Jacobian of this transformation is
\[
\exp \{ \Sigma^J_k [{\hat \sigma}]\} \ \equiv \ ({\rm det}{\tilde \Omega} {\rm det}{\tilde \Omega}^* )^J_k
\]
\[
\ =\ ({\rm det}_M {\hat \Omega}^*_M )^N \ ({\rm det}_{M-1} {\hat \Omega}^{-1}_{M-1} )^{N+1} \ ({\rm det}_{M-1} {\hat \Omega}^{*-1}_{M-1} )^{N+1} \ ({\rm det}_M {\hat \Omega_M} )^N
\]
\be
\ =\ ({\rm det}_M {\hat \Omega}^*_M {\hat \Omega}_M )^N \ ({\rm det}_{M-1}
{\hat \Omega}^{-1}_{M-1} {\hat \Omega}^{*-1}_{M-1} )^{N+1} \ .
\ee
This is an exact formula for the chiral anomaly in the noncommutative case. Obviously, it is gauge invariant.

{\it Note 1}: Replacing ${\hat \Omega}$ in the transformation (60) by an arbitrary unitary operator from ${\hat{\cal H}}_0$ the corresponding determinant (71) will be equal to one. This explicitely indicates, that the pure gauge factors $e^{\pm i{\hat{\lambda}}}$ can be absorbed into spinor fields.    

{\it Note 2}: The formula for the chiral anomaly does not change if one adds gauge invariant terms to the action (68), e.g.
\[
\frac{m}{2J} s{\rm Tr}_N [{\hat{\bar \Psi}}{\hat \Psi}]\ +\ \frac{1}{2J} s{\rm Tr}_N [\mu ({\hat{\bar\Psi}}{\hat\Psi})^2 +\nu ({\hat{\bar\Psi}}\Gamma_0 {\hat\Psi})^2 ]\ .
\]
The first term can be interpreted as a mass term for the field ${\hat \Psi}$, the second one a particular 4-fermionic interaction. Performing the spinor field transformation (60) the fermionic fields ${\hat{\bar \Psi}}_0$, ${\hat \Psi}_0$ does not separate from gauge degrees of freedom: an integration over fermionic gives an additional contribution to the effective gauge field action besides  fermionic determinant (71). 

\subsection{Product formula and commutative limit}
To find the commutative limit we shall represent the operators from ${\hat{\cal H}}_0$ in the coherent state basis (for details see \cite{GP1}). The formula for the coherent states $|\vec{x};N\rangle
\in {\cal F}_N$ reads
\[
|\vec{x};N\rangle \ \equiv |\xi (\vec{x});N\rangle \ \ :=\ \frac{1}{\sqrt{N}}(\chi^+ \xi (\vec{x}))^N |0\rangle
\]
\be 
\ =\ \sum_{n=0}^N \sqrt{\frac{N!}{n!(N-n)!}} \xi^n_1 \vec{x} \xi^{N-n}_2 \vec{x} |n,N-n \rangle \ .
\ee
Here $\vec{x} =\xi^+ (\vec{x})\vec{\sigma} \xi (\vec{x})\in S^2$, $\xi (\vec{x})$ is an arbitrary section of the bundle $S^3 \to S^2$ normalized by $\xi^+ (\vec{x})\xi (\vec{x})=1$. For various $\xi (\vec{x})$ the vectors $|\xi (\vec{x});N\rangle$ differ just by a phase factor. Let $T_N (\gamma )$, $\gamma \in SU(2)$, is the $SU(2)$ group representation of in ${\cal F}_N$ corresponding to the canonical realization (10). Putting $\gamma =C+i\vec{S}. \vec{\sigma}$, $C^2 +{\vec{S}}^2 =1$, it can be shown that

\be
w_N (\gamma ,\vec{x})\ :=\  \langle \vec{x};N|T_N (\gamma )|\vec{x};N
\rangle \ =\ (C+i\vec{S}.\vec{x})^N \ ,
\ee
(for $\vec{S}=(0,0,S)$ the proof is straightforward, eq. (73) follows by rotational invariance). 

To any operator ${\hat f}\in {\hat{\cal H}}_0$ we assign the function $f_N (\vec{x})=\langle \vec{x};N|{\hat f}|\vec{x};N\rangle$. The normalized trace of the operator ${\hat f}_N ={\hat f}|_{{\cal F}_N}$ in the coherent state basis can be expressed as
\be
\frac{1}{N+1} {\rm Tr}_N {\hat f}_N \ =\ \int d\mu f_N (\vec{x}) \ .
\ee
The $*$-product of two functions $f_N (\vec{x})=\langle \vec{x} ;N |{\hat f}|\vec{x} ;N \rangle$ and $g_N (x)=\langle \vec{x}|{\hat g} |\vec{x}\rangle$ is defined by
\be
(f_N *g_N )(\vec{x})\ =\ \langle \vec{x};N|{\hat f}{\hat g}|\vec{x};N\rangle \ .
\ee

Our aim is to express this $*$-product directly in terms of $f_N (\vec{x})$ and $g_N (\vec{x})$ and their derivatives. To achieve this let us express the operator ${\hat f}_N$ as ${\hat f}_N =\int d\gamma T_N (\gamma ){\tilde f}(\gamma))$, $d\gamma$ - Haar measure. Then $f_N (\vec{x})=\int d\gamma {\tilde f}(\gamma ) w_N (\gamma ,\vec{x})$, analogously, $g_N (\vec{x})=\int d\gamma {\tilde g} (\gamma ) w_N (\gamma ,\vec{x})$. From the relation $T_N (\gamma )T_N (\gamma' )=T_N (\gamma \gamma' )$ and the defin

ition of the $*$-product it follows straightforwardly
\[
(f_N *g_N )(\vec{x})\ =\ \int d\gamma d\gamma' {\tilde f}(\gamma ){\tilde g} (\gamma' ) w_N (\gamma \gamma' ,\vec{x}) \ .
\]
Putting $\gamma =C+i\vec{S}.\vec{\sigma}$ and $\gamma'=C'+i\vec{S}'.\vec{\sigma}$ we
obtain
\[
\frac{w_N (\gamma \gamma',\vec{x})}{w_N (\gamma ,\vec{x})w_N (\gamma',\vec{x})}\ =\ \sum_{k=0}^N \frac{N!}{k!(N-k)!}\left[ \frac{i(\vec{S}\times \vec{S}').\vec{x} -(\vec{S}\times\vec{x}).(\vec{S}'\times \vec{x})}{(C+i\vec{S}.\vec{x}) (C'+i\vec{S}'.\vec{x})}\right]^k
\]
\[
\ =\ \sum_{k=0}^N \frac{(N-k)!}{k!N!} \omega_{i_1 j_1} \dots \omega_{i_k j_k} (\partial_{i_1} \dots \partial_{i_k} w_N )(\gamma ,\vec{x})(\partial_{j_1} \dots \partial_{j_k} w_N )(\gamma' ,\vec{x}) \ ,
\]
where $\omega_{ij} =i\varepsilon_{ijk} x_k -x_i x_j +\delta_{ij}$. This
induces the desired explicit $*$-product formula
\be
(f_N *g_N )(\vec{x})= \sum_{k=0}^N \frac{(N-k)!}{k!N!} \omega_{i_1 j_1}
\dots \omega_{i_k j_k}(\partial_{i_1} \dots \partial_{i_k} f_N )(\vec{x})\,
(\partial_{j_1} \dots \partial_{j_k} g_N )(\vec{x}) \ .
\ee

From (76) it follows the asymptotic formula for the $k$-th $*$-power 
\[
(f_N *\dots *f_N )(\vec{x}) = f^k_N (\vec{x})\ +\ \frac{k(k-1)}{2N} f^{k-2}_N (\vec{x}) (X_i f_N (\vec{x}))^2 +o(N^{-2})\ .
\]
If $F(z)$ is a polynomial (analytic function) this allows us to obtain the asymptotic formula linking $F_N (f)(x)$ to $f_N (x)$:
\be
F_N(f)(\vec{x}) \ =\ F(f_N (\vec{x}))\ +\ \frac{1}{2N}F''(f_N (\vec{x}))\, (X_i f_N (\vec{x}))^2 +o(N^{-2})\ .
\ee
This is the key formula we use for the calculation of the commutative limit of the chiral anomaly. 

Applying it to the function $\Omega \Omega^* =\exp [2e_o \sigma ]$ we obtain
\[
(\Omega \Omega^* )_N (\vec{x})\ =\ e^{2e_o \sigma_N (\vec{x})} \ +\ \frac{2e^2_o}{N} e^{2e_o \sigma_N (\vec{x})} (X_i \sigma_N (\vec{x}))^2 \ +\  o(e^2_o N^{-2} )\ ,
\]
or,
\be
\ln (\Omega \Omega^* )_N (\vec{x})\ =\ 2e_o \sigma_N (\vec{x}) \ +\
\frac{2e^2_o}{N} (X_i \sigma_N (\vec{x}))^2 \ +\ o(e^2_o N^{-2} )\ .
\ee
This gives the asymptotic formula for the determinant
\[
{\rm det}_N (\Omega \Omega^* )_N \ =\ \exp [{\rm Tr}_N \ln (\Omega
\Omega^* )_N ]
\]
\be
\ =\ \exp \{ 2e^2_o \int d\mu (X_i \sigma_N (x))^2 ]\ +\ o(e^2_o N^{-1} )\} \ ,
\ee
(the linear term in $\sigma_N (\vec{x})$ in (78) does not contribute to the integral). Using (79) for both factors in the chiral anomaly formula (71) and renormalizing the constant $e_o$ properly by $2J e^2_o =q^2_o$, we obtain in the commutative limit ($J\to\infty$, $k$ - fixed) the quantum correction $\Sigma^J_k [\sigma ]$ to the bosonic action
\be
\Sigma^J_k [\sigma ]\ =\ 2q_o^2 \int d\mu (X_i \sigma (x))^2
\ +\ o(J^{-1} ) \ .
\ee
Here we have used relation $\int d\mu (X_i \sigma_N (x))^2 =\int d\mu (X_i \sigma (x))^2 +o(N^{-3} )$ which is valid provided that the commutative prepotential $\sigma (\vec{x})$ leads to a finite contribution to the commutative gauge field field action (59). Eq. (80) reproduces the standard commutative result (66).

{\it Note}: Let us we replace the factorials in (76) by $\Gamma$-functions and $N$ in the arguments by a real parameter $\varepsilon^{-1}$. Performing a power expansion in $\varepsilon$ we obtain, for generic $\varepsilon$, a divergent but Borel summable series corresponding to the Kontsevich quantization formula, \cite{Kon}. For the "critical" values $\varepsilon = N^{-1}$ the formal power series quantization is unitarizable and reduces to the one described by (76).  

\section{Concluding remarks}
The essential steps in our approach to the fuzzy Schwinger model can be summarized as follows:

i) In order to guarantee the gauge invariance we have to work with all (fuzzy) canonical realizations of fields (i.e. fields are realized as operators in the Fock space and not as some finite dimensional matrices).

ii) We have been able to define the rotationally invariant field action containing only finite number of modes of dynamical fields, i.e. with respect to the dynamical modes the resulting model is finite dimensional (matrix) model. Such mode restriction is not possible for the pure gauge prepotential $\lambda$.

iii) The dynamical modes were quantized within functional integral approach, and $\lambda$ appears as a background field. However, the quantum mean values of gauge invariant functionals do not depend on a particular choice of $\lambda$, it can be fixed arbitrarily (e.g. $\lambda =0$). The resulting model for dymamical modes is a finite dimensional matrix model, and consequently is nonperturbatively UV-regular. 

iv) In a simple and direct way we derived an exact formula for the chiral anomaly. Using the explicit formula for the $*$-product on a sphere we recovered in the commutative limit the standard reasult.

It would be desirable to generalize the model to 4D case. This will be not straightforward, since various steps in our construction are linked with the particular properties of 2D sphere. However, one can expect that some specific features will survive in 4D.
\vspace{1cm}

{\bf Acknowledgement}: The author would like to thank to Harald Grosse (Vienna) for many valuable and fruitfull discussions. This work was supported by the "Fond zur F\"{o}rderung der Wissenschaftlichen Forschung in \"{O}sterreich" project P11783-PHY and by the Slovak Grant Agency VEGA project 1/4305/97.

\end{document}